\documentclass[preprint,floatfix,nofootinbib]{revtex4-1}
\usepackage{xcolor} 
\usepackage{changes} 
\usepackage{hyperref} 
\usepackage{url} 
\usepackage{float} 
\usepackage{caption} 
\usepackage{tabularx} 
\usepackage{graphicx} 
\usepackage[utf8]{inputenc}
\usepackage[T1]{fontenc}
\usepackage{mhchem}
\usepackage{mwe} 
\usepackage{subfig} 
\usepackage{silence}
\begin{document} 

\title{The Nature of Electrophilic Oxygen : Insights from Periodic Density Functional Theory Investigations}

\author{Nivedita Kenge}
\affiliation{Physical and Materials Chemistry Division, CSIR-National Chemical
Laboratory, Pune, India - 411008}
\author{Sameer Pitale, and Kavita Joshi}
\email{k.joshi@ncl.res.in,kavita.p.joshi@gmail.com} 
\affiliation{Physical and Materials Chemistry Division, CSIR-National Chemical
Laboratory, Pune, India - 411008}
\affiliation{Academy of Scientific and Innovative Research (AcSIR),
Anusandhan Bhawan, 2, Rafi Marg, New Delhi – 110 001, India}


\begin{abstract}
Increasing demand of ethylene oxide and the cost of versatile chemical ethene has been a 
driving force for understanding mechanism of epoxidation to develop highly selective catalytic 
process. Direct epoxidation
is a proposed mechanism which in theory provides 100\% selectivity. A key aspect of this 
mechanism is an electrophilic oxygen (O$_{ele}$) species forming on the Ag surface. In the past two and half 
decades, large number of theoretical and experimental investigations have tried to elucidate
formation of O$_{ele}$ on Ag surface with little success. Equipped with this rich literature 
on Ag-O interactions, we investigate the same using periodic DFT calculations to further 
understand how silver surface and oxygen interact with each other from a chemical standpoint. 
Based on energetics, L\"{o}wdin charges, topologies and {\it pdos} data described in this study, we scrutinize
the established notions of O$_{ele}$. Our study provides no evidence in support of O$_{ele}$ being 
an atomic species nor a diatomic molecular species. In fact, a triatomic molecular species described in this work
bears multiple signatures which are very convincing evidence for considering it
as the most sought for electrophilic entity. 
\end{abstract} 

\maketitle 

\section{Introduction} 
Silver catalyzed ethene (En) epoxidation by oxygen to produce ethylene oxide (EtO) is one 
of the most well-known kinetically controlled reaction in chemistry. It is also one of the 
most studied example in discipline of heterogeneous catalysis, owing to the great economic 
importance of EtO as well as En.\cite{sachtler_mechanism_1981, chemical_weekly, ozbek_mechanism_2013}
Ethene in presence of oxygen undergoes total 
combustion to CO$_{2}$ and H$_{2}$O virtually under all conditions except when the reaction 
is carried out in presence of Ag, where (for unpromoted Ag) the product is 50\% EtO and 
rest are the combustion products.\cite{1935patent}
From thermodynamic data, it is very clear that combustion products are far more stable owing to their reduced free energy.
Even EtO, very rapidly decomposes into same combustion products due to its thermochemical instability.
EtO can thus form only under conditions of kinetic control as it has low free energy of activation and not at conditions of 
thermodynamic equilibrium.\cite{thesis}
A kinetically controlled reaction involves reaction intermediate with low free energy of activation and is independent of free energy of formation of the product.\cite{March_organic}
At present, 50\% selectivity of bare Ag towards epoxidation is well understood through Oxometallacycle 
(OMC) intermediate based reaction mechanism proposed by Linic and Barteau.\cite{linic2002JACS} 
Chief 
triumph of OMC mechanism is that it explains all the experimental observations not well 
understood over the decades. This includes proving the hypothesis of common 
intermediate for EtO and combustion products,\cite{bell1975} 
C-H bond breaking and proton transfer as rate 
limiting step for combustion, and 
excellent correlation to micro kinetic data.\cite{backx_reactivity_1981,van_santen_mechanism_1987,linic_construction_2003} Still 
OMC cannot explain $\sim 90\%$ selectivity of modern industrial catalysts.\cite{2014patent,ozbek_mechanism_2013} 
In order to account for selectivity of industrial catalyst, it has been hypothesized that reaction under 
those conditions must go through direct epoxidation 
wherein an electrophilic oxygen 
(O$_{ele}$) species is formed on the promoted Ag surface under industrial reaction 
conditions.\cite{van_santen_mechanism_1987} 
EtO is thus formed as a result of an electrophilic attack on ethene through an 
addition reaction. Theoretically, such a mechanism has 100\% selectivity for EtO and thus 
provides very lucrative basis for further improvement of industrial processes.
Though theoretical 
aspects of direct epoxidation mechanism are fairly straight forward,\cite{twigg_1942} identity 
of O$_{ele}$ has not been established unanimously to date.
Nearly all the research on epoxidation 
in $21^{st}$ century and most from the f\mbox{}inal decade of the last century is focused on 
characterization of O$_{ele}$.\cite{avdeev_quasimolecular_2000, bao_interaction_1996,bukhtiyarov_combined_2000,
Bukhtiyarov2001, li_oxygen_nodate, kaichev_nature_2003, schnadt_experimental_2009,
fellah_epoxidation_2011, rocha_silveroxygen_2012, bocklein_high-pressure_2013, jones_insights_2015} 
And though considerable work has been already done in this 
f\mbox{}ield, the exact nature, electronic conf\mbox{}iguration or topological data about O$_{ele}$ has 
remained controversial.\cite{bocklein_high-pressure_2013, jones_insights_2015, martin_2014} 
Until 2001, when Bukhtiyarov {\it et al.} published XANES, XPS, and UPS studies on Ag-O 
interactions, concluding O$_{ele}$ has to be atomic in nature, whether O$_{ele}$ is atomic 
or molecular was also the cause of considerable debate.\cite{Bukhtiyarov2001,avdeev_quasimolecular_2000,li_oxygen_nodate} 

Density Functional Theory (DFT) based studies have proven themselves as 
indispensable tools for surface chemistry.
OMC mechanism was also discovered through DFT 
studies.\cite{linic2002JACS} Since then DFT based investigations of Ag-O interaction has become an 
attractive avenue for computational research.
Through extensive computational studies carried out on multiple Ag 
facets, no species could be isolated which can f\mbox{}it the description of O$_{ele}$.\cite{Bukhtiyarov2001, 
li_oxygen_nodate,kaichev_nature_2003,rocha_silveroxygen_2012,jones_insights_2015,bocklein_high-pressure_2013,schnadt_experimental_2009,fellah_epoxidation_2011} But in 
wealth of thermodynamic and physical data available through these studies, there seems to 
be glaring omission of chemical aspects of Ag-O interactions. Ag-O interactions are 
surprisingly diverse in nature and highly dependent of temperature, pressure as well as Ag 
facets and concentration of oxygen.\cite{bocquet_new_nodate,schnadt_experimental_2009, 
derouin_thermally_2016} 

Oxidation of Ag surface to form bulk oxide like Ag$_{2}$O is
a complicated process as observed in the recent studies.\cite{Gopinath2018, 
Morgenstern2016, Keffer2017, Friend2018}
Thin bulk oxide like Ag$_{2}$O layer formation shows reversible behavior
with increase in temperature.\cite{Gopinath2018}
Recent experimental studies on Ag clusters supported on
silica reveal that, larger clusters with 55 Ag atoms do not undergo complete oxidation.\cite{Lunskens2017} 
Instead of forming stable Ag$_{2}$O
phase, Ag prefers to maximize its coordination with O through interaction with molecular oxygen 
and this results in observed propensity of atomic O to occupy subsurface sites.\cite{Keffer2017} 
Further, O$_{2}$ chemisorption results in covalent interactions with Ag atoms.\cite{Morgenstern2016}
Combined DFT and experimental investigations revealed that a low coverage of adsorbed atomic oxygen 
exists on Ag surface.\cite{Piccinin2018} Further, studies carried out by Jones {\it et.al}
concluded that O-1s signature of O$_{ele}$ can only come
from covalently bonded oxygen.\cite{jones_insights_2015} Additionally, authors noted that
surface adsorbed atomic oxygen do not interact covalently enough to
give rise to O$_{ele}$ signatures in O-1s binding energies. They proposed that
further research should focus on understanding how to make Ag-O
interactions more covalent. In this light, recent publication on Ag-O interactions described the formation 
of SO$_{4}$
on the surface of Ag due to
sulphurous impurities as a suspected source of O$_{ele}$.\cite{Piccinin2018(2)}

It is worth noting here some industrial facts regarding epoxidation.
For epoxidation process at an industrial scale, apart from Ag catalysts, high pressure (5-30 bar) and 
controlled temperature in the range of (423-623 K) are necessary conditions.\cite{2014patent} Though, the 
reported selectivities at industrial level are $\sim 90\%$, 
Et to EtO conversion rates are typically low around 15\%.\cite{2014patent,engineering_chem_research} A
recent experimental study based on temporal analysis at atmospheric pressure, of epoxidation products, demonstrates 
that conversion of Et to EtO never exceeded 20\% at 44\% selectivity typical for bare and unpromoted Ag.\cite{Scharfenberg2017} 
Thus, high pressure, high oxygen concentration (compared to Et) in the feed and strict temperature control are unique 
characteristics of industrial scale ethene epoxidation apart from promoted catalytic surface.\par
\setlength{\parskip}{0pt}
To search for O$_{ele}$ which is a key to direct epoxidation, we have modeled Ag-O interactions 
at various monolayer concentrations of O on Ag(100) surface. 
For an electrophilic oxygen to exist on Ag surface, it should have following two characteristic chemical features.
Firstly, it should have a positive charge and secondly, it should be a covalently bonded 
species with empty states in the energy range 0-6 eV.\cite{kokalj_2002,jones_insights_2015} 
These empty states enable such species to accept electrons from ethene $\pi$ orbitals. 
From our review of the literature, we could not f\mbox{}ind any work
describing existence of such empty states.
Therefore, we reinvestigate Ag-O interactions to uncover
the nature of bonding as well as conf\mbox{}igurations leading to O$_{ele}$ signatures.
Through exhaustive modeling of topologies using DFT, partial density of states ({\it pdos})
and L\"{o}wdin charge calculations we present a chemically intuitive picture of Ag-O
interactions to seek answer for questions such as, what is the nature of Ag-O
interactions? Are they predominantly ionic or covalent? Do electrophilic and nucleophilic oxygen form
from Ag-O interactions alone? Is it possible to observe empty states for oxygen
species chemisorbed on Ag surface? Could such species posses positive charge?
We hope our ef\mbox{}forts
could provide some insight which can help in pursuit of elucidating the direct
epoxidation mechanism.

\section{Computational Details}

We have investigated interaction of O with Ag(100) surface by employing Kohn-Sham 
formulation of DFT. The interaction between electrons and ions 
was modeled with Projector Augmented Wave pseudopotential with Perdew-Burke-Ernzerhof (PBE) 
exchange-correlation within generalized gradient 
approximation, as implemented in the plane wave code, Quantum ESPRESSO.\cite{paw1,paw2,pbe1,pbe2,qe} 
Energy cutof\mbox{}f for plane-waves was kept at 70 Ry and 700 Ry for charge density. 
The Ag surface was modeled by cleaving a surface with 4 layers in 
(100) direction. The vacuum along $z$-axis which is also (100) direction of the 
crystal was varied from 10 {\AA} till 30 {\AA} with the step of 2 {\AA}. It was found that 20 
{\AA} of vacuum is suf\mbox{}f\mbox{}icient to avoid interaction between adjacent images of planes along 
the $z$-direction. Since adsorption of reactant molecule/s on one side of slab gives rise 
to inhomogeneity in electric f\mbox{}ield, dipole f\mbox{}ield correction was applied in 
the $z$-direction in order to compensate this inhomogeneity. Geometry optimization 
was carried out with a force cutof\mbox{f} of 10$^{-3}$ a.u. on the unf\mbox{i}xed atoms and 
the total energies were converged below 10$^{-4}$ a.u.  A Monkhorst-Pack grid of 12x12x1 was 
used which resulted into 72 \texttt{k}-points in IBZ to emulate the solid slab.

As we focused on silver-oxygen interaction, the system was chosen in such a way that it can
be exhaustively studied at optimal computational expenses. In a $2$X$2$ slab, based on
symmetry arguments, three distinct surface sites exist.
Also the calculations are carried out under periodic boundary conditions and thus the system
can be repeated inf\mbox{}initely to model macroscopic surface. But as unit cell for such
macroscopic model consists of surface created by four Ag atoms, inclusion of single oxygen
atom on surface is equivalent to 1/4 monolayer (ML) coverage of oxygen
which is equivalent to the model of pristine silver surface coated with 1/4
ML of oxygen under experimental conditions. Under these circumstances, our model enables us
to study $0.25$ ML (one oxygen), $0.5$ ML (two oxygen atoms), $0.75$ ML (three oxygen
atoms) and 1 ML (four oxygen atoms). As attaining 1 ML is very dif\mbox{}f\mbox{}icult under experimental
conditions and adding four atoms in such a small system results in severe distortions, we
have not included cases of the same in this study. We have considered all possible
conf\mbox{}igurations under $0.25$, $0.5$, $0.75$ monolayers, where oxygen atom or molecule can
occupy unique positions. Thus, it is ensured that all types of silver-oxygen
bonding conf\mbox{}igurations are investigated. A small modeled system has obvious
disadvantages when it comes to elucidation of reaction intermediates and mechanisms. However,
we have def\mbox{}ined the goal of this study, to investigate adsorption of oxygen on silver and
to understand the nature of bonding in such interactions. Hence, this small system provides a modest
starting point. In the following section, we discuss the 
interaction between O species and Ag(100) surface as a function of increasing concentration 
of O.  As noted earlier, the surface was modeled by considering four layers of Ag(100) 
with 2x2 supercell.  The bottom most layer is f\mbox{}ixed. Thus, our simulation box contains 16 Ag atoms. 
Number of oxygen atoms varies based on the monolayer concentration of O.

\section{Results and Discussion}
\begingroup
\setlength{\tabcolsep}{8pt}
\renewcommand{\arraystretch}{0.9}
\begin{table}[H] 
\centering
\begin{tabular}{c c c c c}
Conf\mbox{}iguration & $\Delta$E      & Ag-O          & Ag-O            & L\"{o}wdin          \\ 
Description   &   (eV)         & coordination  & bond length \AA  & charge on O     \\ 
\hline
O$_{4FH}$     & 0.00           & 4             &  2.25           & 6.53 \\ 
O$_{2FB}$     & 0.82           & 2             &  2.05           & 6.46 \\ 
O$_{1FT}$     & 1.89           & 1             &  1.94           & 6.39 \\ 
\hline
\end{tabular} 
\caption{\small {Relative energy ($\Delta$E), Ag-O coordination along with bond length, 
and L\"{o}wdin charges on oxygen for dif\mbox{}ferent positions of oxygen at 0.25 ML coverage. The most stable 
structure has highly coordinated oxygen with maximum L\"{o}wdin charge.}}
\label{tab1}
\end{table}
\endgroup

\begin{figure}[h]
    \includegraphics[width=0.5\textwidth]{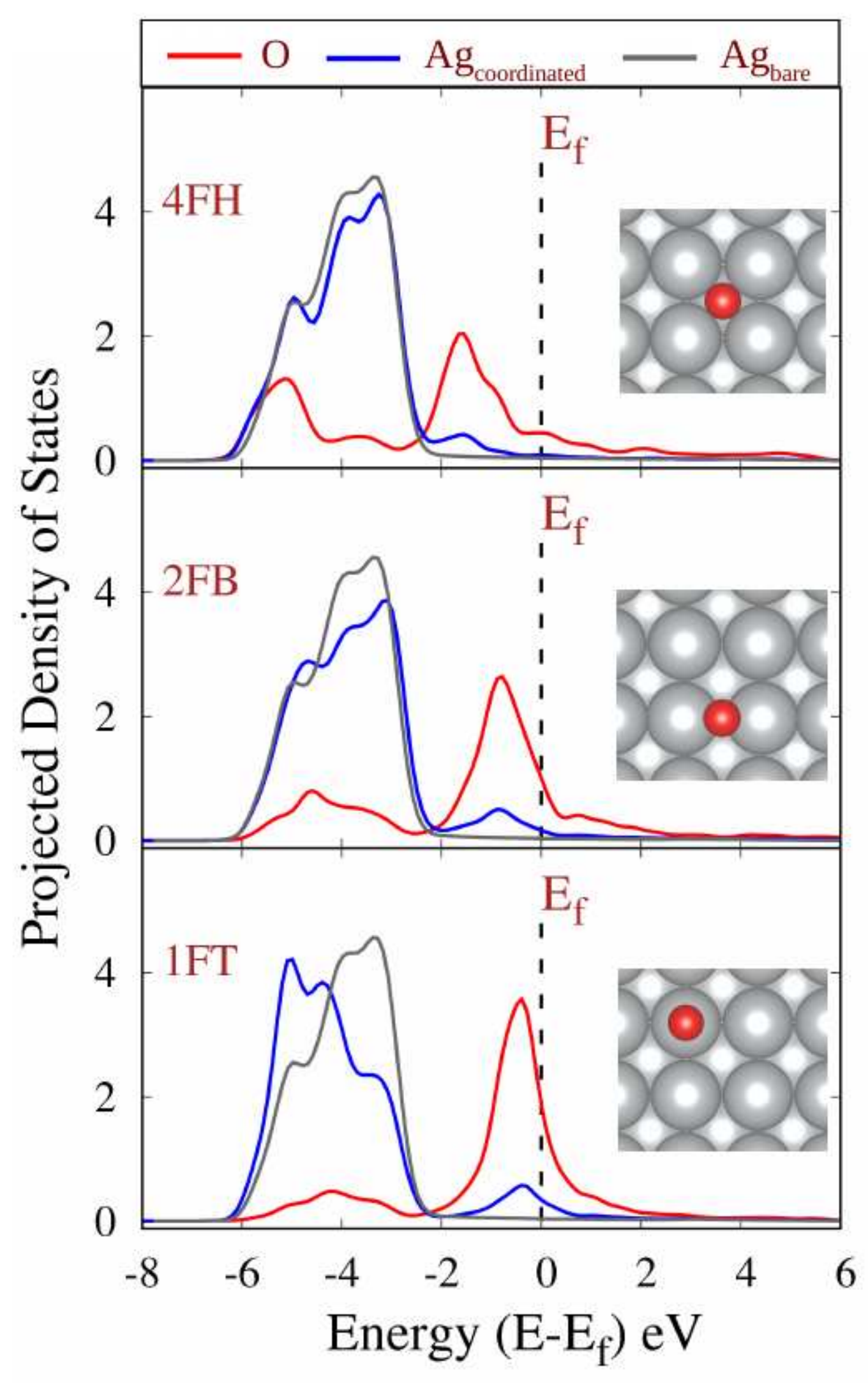}
  \caption{\small {Site dependent {\it pdos} for 4d-states of surface Ag atoms in bare slab (gray), 
4d-states of surface Ag atoms coordinated with O (blue) 
and 2p-states of adsorbed O on Ag surface (red). Their respective configurations are also shown.
2p-states for four fold coordinated oxygen atom are the most delocalized among all the configurations .}}
  \label{fig1}
\end{figure}

In this section, we take a closer look in the chemical signatures 
and trends generated therein to relate the data with observations 
made both experimentally and theoretically in prior literature. 
The focus of this discussion is to bring out probable conf\mbox{}igurations which will 
support existence of electrophilic oxygen species which leads to direct epoxidation.
For the case of 0.25 ML, we have studied interactions of oxygen with Ag surface at three 
distinct surface sites such as 4 fold hollow (4FH), 2 fold bridge (2FB), and 1 fold top (1FT).
Details regarding the energetics, Ag-O bond lengths as well as coordination, and  L\"{o}wdin charges of
oxygen atom are summarized  in Table\ \ref{tab1}. The conf\mbox{}igurations along with site dependent {\it pdos} are 
shown in Fig.\ \ref{fig1}.
 
Out of these three surface conf\mbox{}igurations, oxygen adsorbed at hollow site (4FH) is energetically 
the most pref\mbox{}fered conf\mbox{}iguration followed by the bridge position (2FB) and the least stable 
is on-top site (1FT). 
Ag-O bond length increases with coordination and is noted
in Table\ \ref{tab1}. L\"owdin charge analysis manifests that charge of higher coordinated oxygen species 
is greater in magnitude  while for lower coordinated oxygen atom,
 magnitude of charge is lesser. 
Site dependent {\it pdos} brings out interesting observations related to the site specific 
interaction between surface Ag atoms and adsorbed oxygen as shown in Fig.\ \ref{fig1}. The {\it pdos} for 
O-$2p$ associated with on-top oxygen peaks near Fermi level (E$_f$) and it is relatively sharper. 
On the other hand, the peak position of O-$2p$ shifts away 
from E$_f$ and is accompanied with broadening of peaks as coordination of adsorbed oxygen increases 
for 2FB and 4FH conf\mbox{}igurations. When considering 2$p$-4$d$ overlap, a sharp peak in {\it pdos} 
represents localized electron distribution and thus ionic character of the bond. Conversely, broader peak 
suggests delocalized state and alludes to covalent character of the bond.\cite{Hoffmann88} 

Surface Ag-O interactions, modeled on the basis of Ag$_2$O in previous investigations, considered it to be 
ionic in nature.\cite{bukhtiyarov_combined_2006} Contrary to this, 
our studies bring out that surface Ag-O interactions are preferably covalent in nature with little tendency 
towards formation of Ag$_2$O. This preferred covalent bonding is unique 
for Ag among other transition metals where oxidation of surface and formation of ions is a norm. 
Because of high energetic cost of ion formation, covalent overlap with multiple states seems to 
be preferred mode of bonding for oxygen on Ag surface. This explains, why hollow position 
is a preferred site over bridge and top positions. 
Another important observation provided by {\it pdos} is a lack of empty states 
near E$_{f}$. Availability of these states near E$_{f}$ is an essential criterion for reaction to occur. 
As none of the atomic oxygen species possess any compatible states above E$_{f}$ for accepting $\pi$-electrons from 
ethene, we conclude that single atomic oxygen species should be precluded as contributing 
factor in the direct epoxidation.
\begin{figure}[h]
  \centering
  \begin{minipage}[b]{0.45\linewidth}
   \centering 
   \subfloat[]{\label{fig2:a}\includegraphics[width=\textwidth]{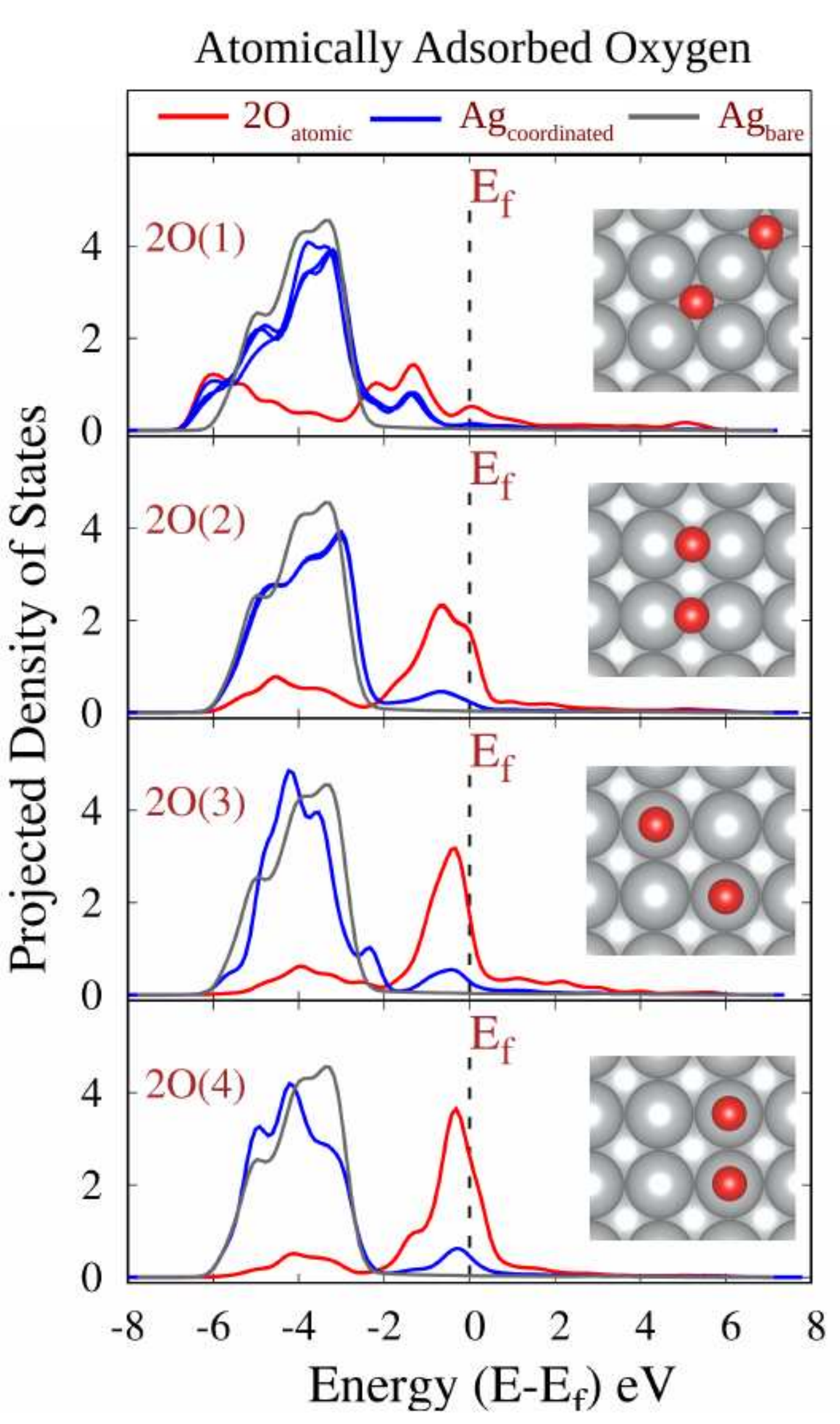}}
  \end{minipage}
  \hfill
  \begin{minipage}[b]{0.45\linewidth}
   \centering
   \subfloat[]{\label{fig2:b}\includegraphics[width=\textwidth]{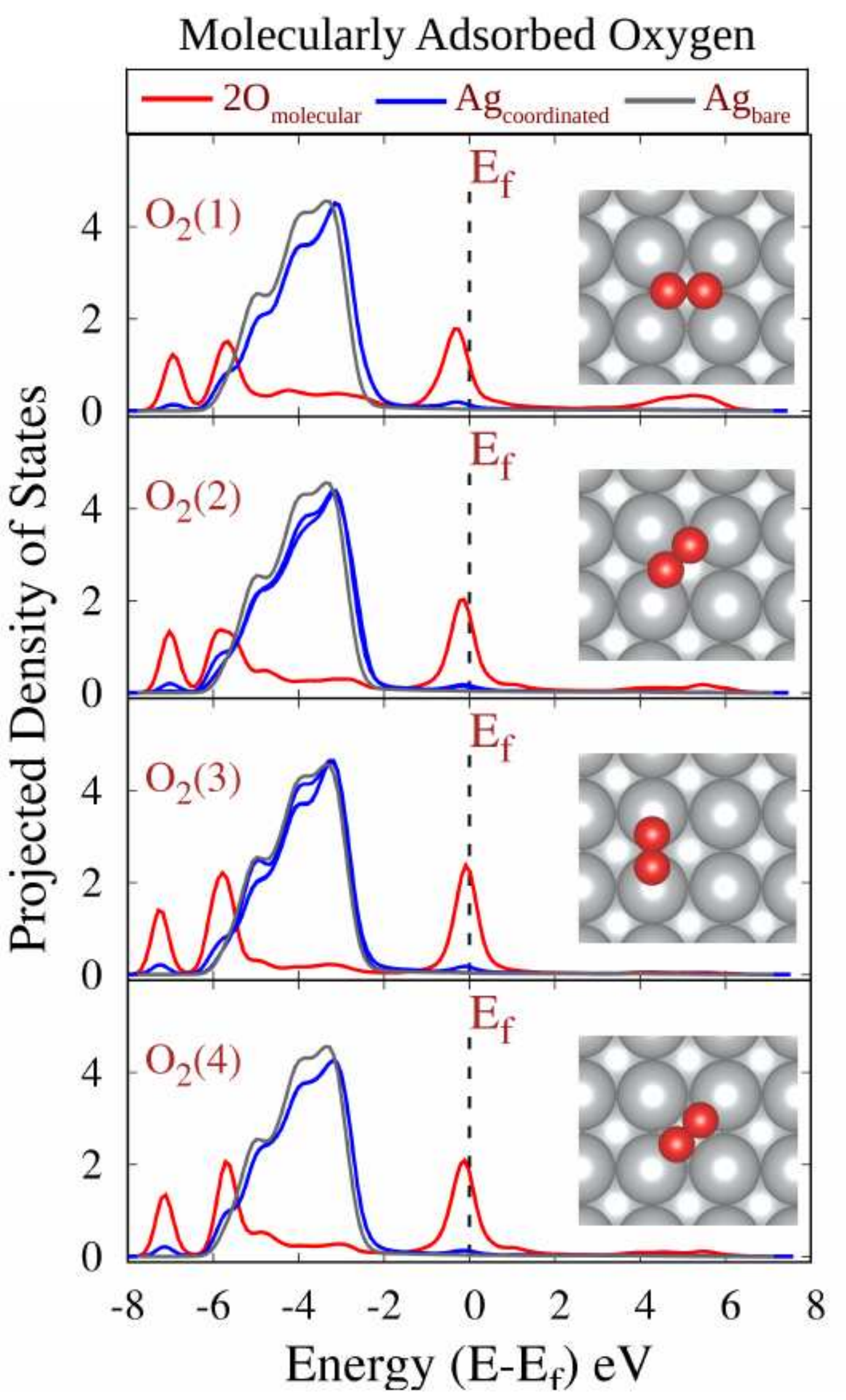}}
  \end{minipage}
  \caption{\small {Site dependent {\it pdos} for (a) atomically adsorbed oxygen (2O) and (b) molecularly adsorbed oxygen (O$_{2}$)
   on Ag(100) surface at 0.5 ML along with their respective configurations. Ag-4d states of bare slab are shown in gray,
   4d states of coordinated Ag surface atoms are shown in blue and 2p states of oxygen are indicated in red color. 
   {\it pdos} for atomically adsorbed oxygen shows striking similarity with that of single oxygen case discussed earlier.}}
  \label{fig2}
\end{figure}

Considering $0.5$ ML coverage, all the conf\mbox{}igurations tested can be classif\mbox{}ied into two groups, 
viz. atomically adsorbed oxygen atoms (2O)
distributed over three distinct sites and molecularly adsorbed oxygen (O$_2$) placed at various sites.
Adsorption of two atomic oxygen leads to eight initial conf\mbox{}igurations 
and adsorption of molecular oxygen leads to f\mbox{}ive initial conf\mbox{}igurations. 
These thirteen initial conf\mbox{}igurations resulted into eight stable conf\mbox{}igurations upon relaxation
and are shown in Fig.\ {\ref{fig2}} along with their site dependent {\it pdos}.
Cases in which two oxygen atoms are more than 2 {\AA} apart are
considered as atomically adsorbed.
For these eight stable conf\mbox{}igurations, relative energies,
Ag-O coordination, Ag-O and O-O bond lengths, and L\"owdin charges are summarized in Table\ \ref{tab2} where 2O 
conf\mbox{}igurations are shown in red color and O$_{2}$ conf\mbox{}igurations are shown in green.\\ 
\begingroup
\setlength{\tabcolsep}{8pt}
\renewcommand{\arraystretch}{0.9}
\begin{table}[h!]
\centering
\begin{tabular}{c c c c c c c}
Sr. No. & Conf\mbox{}iguration        & $\Delta$E    & Ag-O            & Ag-O             & O-O               & L\"owdin Charge \\
        & Description          &        (eV)  & coordination    & bond length(\AA)  & bond length(\AA)  & on O          \\
\hline
1. & \textcolor{red}{2O(1)} &\textcolor{red}{0.00} &\textcolor{red}{4} &\textcolor{red}{2.11-2.21} &\textcolor{red}{-} &\textcolor{red}{6.49} \\

2. & \textcolor{green!30!black}{O$_{2}$(1)} &\textcolor{green!30!black}{0.73} &\textcolor{green!30!black}{2} &\textcolor{green!30!black}{2.26} &\textcolor{green!30!black}{1.42} &\textcolor{green!30!black}{6.52} \\

3. & \textcolor{green!30!black}{O$_{2}$(2)} &\textcolor{green!30!black}{1.16} &\textcolor{green!30!black}{2} &\textcolor{green!30!black}{2.27-2.41} &\textcolor{green!30!black}{1.37} &\textcolor{green!30!black}{6.18} \\

4. & \textcolor{green!30!black}{O$_{2}$(3)} &\textcolor{green!30!black}{1.17} &\textcolor{green!30!black}{1} &\textcolor{green!30!black}{2.22} &\textcolor{green!30!black}{1.31} &\textcolor{green!30!black}{6.09} \\

5. & \textcolor{green!30!black}{O$_{2}$(4)} &\textcolor{green!30!black}{1.18} &\textcolor{green!30!black}{1} &\textcolor{green!30!black}{2.26} &\textcolor{green!30!black}{1.36} &\textcolor{green!30!black}{6.17} \\

6. & \textcolor{red}{2O(2)} &\textcolor{red}{2.08} &\textcolor{red}{2} &\textcolor{red}{2.04} &\textcolor{red}{-} &\textcolor{red}{6.35} \\

7. & \textcolor{red}{2O(3)} &\textcolor{red}{3.62} &\textcolor{red}{1} &\textcolor{red}{1.97} &\textcolor{red}{-} &\textcolor{red}{6.43} \\

8. & \textcolor{red}{2O(4)} &\textcolor{red}{4.32} &\textcolor{red}{1} &\textcolor{red}{1.97} &\textcolor{red}{-} &\textcolor{red}{6.28} \\

\hline
\end{tabular}
\caption{\small {Relative energy ($\Delta$E), Ag-O coordination along with bond length,
O-O bond lengths, and L\"{o}wdin charges on oxygen for various conf\mbox{}igurations of oxygen at 0.5 ML coverage.
Atomically adsorbed conf\mbox{}igurations (2O(1) to 2O(4)) are shown in red color and molecularly
adsorbed conf\mbox{}igurations are indicated in (O$_{2}$(1) to O$_{2}$(4)) in green color.}}
\label{tab2}
\end{table}
\endgroup
Although, energetically the most stable conf\mbox{}iguration is the adsorption of two atomic oxygen at
hollow site, it is followed by four conf\mbox{}igurations where adsorbed oxygen is in molecular form. 
Interestingly, among these molecularly adsorbed O$_{2}$ conf\mbox{}igurations, 
O-O bond is the weakest in O$_{2}$(1). This signif\mbox{}ies importance of 
Ag-O coordination even for molecularly adsorbed oxygen. 

{\it pdos} along with the configuration for atomically adsorbed oxygens is shown in Fig.\ref{fig2:a} 
where as configurations with molecularly adsorbed oxygen are shown in Fig\ref{fig2:b}.
As seen in Fig.\ \ref{fig2:a}, {\it pdos} for atomic cases are strikingly similar to their 
counterparts in 1O scenario.  Hence, discussion in this section will focus on the molecularly 
adsorbed oxygen. In case of molecularly adsorbed oxygen, {\it pdos} for all the cases are very similar,
 exhibiting signature of molecular oxygen which is considerably dif\mbox{}ferent than that of atomically 
adsorbed oxygen. Further, the most stable molecularly adsorbed oxygen also exhibit more delocalisation 
among this class.  As established for 1O cases, increase in overall delocalisation between O-2p and Ag-4d 
states indicates increasing stability of the conf\mbox{}iguration. Similarly, molecular structure with 
maximum potential for electron delocalisation turns out to be the most stable as seen in case of O$_{2}$ 
at hollow (O$_{2}$(1)) and as emphasized in O$_{2}$ across corner Ag 
(O$_{2}$(2)), where at the cost of steric crowding better Ag-O and O-O bonding is achieved. 
This emphasizes nature of Ag-O interactions which do not favor formation of well def\mbox{}ined molecules 
but weakly bonded moieties with maximum possible coordination for each atom involved.
L\"{o}wdin charges preclude consideration of any oxygen species studied here as an electrophile 
from ionic standpoint. Further, lack of empty states near E$_{f}$ emphasis their inability to accept $\pi$ electrons 
from ethene. Thus, these conf\mbox{}igurations cannot be considered as electrophiles from covalent standpoint also.
Owing to this analysis, such dissociatively adsorbed diatomic molecular oxygen species should be precluded as 
contributing factor in direct epoxidation.

\begin{figure}[h]
  \centering
    \includegraphics[width=0.45\textwidth]{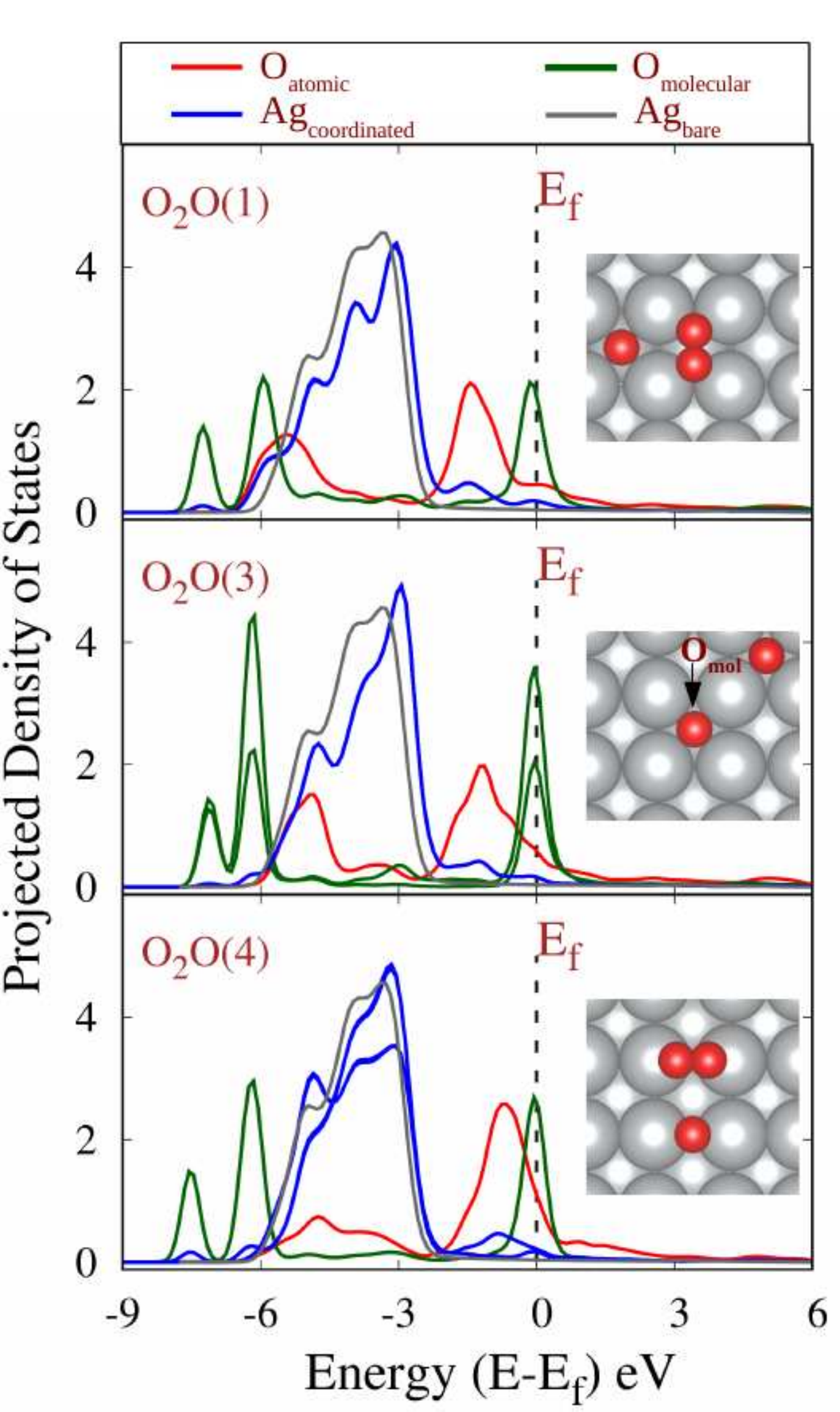}
  \caption{\small {Site specif\mbox{}ic {\it pdos} of O$_{2}$O conf\mbox{}igurations at $0.75$ ML along with their 
  respective configurations. 2p states of O$_{2}$ are shown in green and 2p states of atomic O are shown in red color.
  The pdos for combined atomic and molecularly adsorbed oxygen resembles chemical features of 
  conf\mbox{}igurations at $0.25$ ML and molecular conf\mbox{}igurations at $0.5$ ML.}}
  \label{fig3}
\end{figure}

In case of 0.75 ML, we have modeled the system by adsorbing three atomic oxygen (3O), 
combination of atomic and diatomic molecular oxygen (O$_{2}$O) and triatomic molecular 
oxygen (O$_{3}$) to account for all possible combinations. We would like to emphasis that 
O$_{3}$ is dif\mbox{}ferent than molecular allotropic ozone. In this work, we refer triatomic 
oxygen species as O$_{3}$ and allotrope as ozone.
Three atomic oxygens lead to twelve initial conf\mbox{}igurations, combination of atomic and 
diatomic molecular oxygen leads to eighteen initial conf\mbox{}igurations,
and triatomic molecular oxygen lead to six initial conf\mbox{}igurations. 
All these thirty six initial conf\mbox{}igurations resulted into twelve stable conf\mbox{}igurations 
upon relaxation. These conf\mbox{}igurations are classif\mbox{}ied into two groups viz., 
f\mbox{}ive conf\mbox{}igurations of O$_{2}$O and seven conf\mbox{}igurations of O$_{3}$. Fig.\ \ref{fig3} represents 
three distinct conf\mbox{}igurations out of 
f\mbox{}ive.\footnote{Two conf\mbox{}igurations (O$_{2}$O(2) and O$_{2}$O(5)) 
are not included on the basis of their similar {\it pdos} nature.} 
Energetics, Ag-O and O-O coordination, Ag-O and O-O bond lengths, and L\"owdin charges on oxygen 
of all relaxed conf\mbox{}igurations are summarized in Table\ \ref{tab3} where O$_{2}$O cases 
are shown in blue color and O$_{3}$ cases are shown in green color for a clear distinction. 
Conf\mbox{}igurations which show positive charge on oxygen are highlighted.

\begin{figure}[h]
  \begin{minipage}[b]{0.45\linewidth}
   \centering
    \subfloat[]{\label{fig4:a}\includegraphics[width=\textwidth]{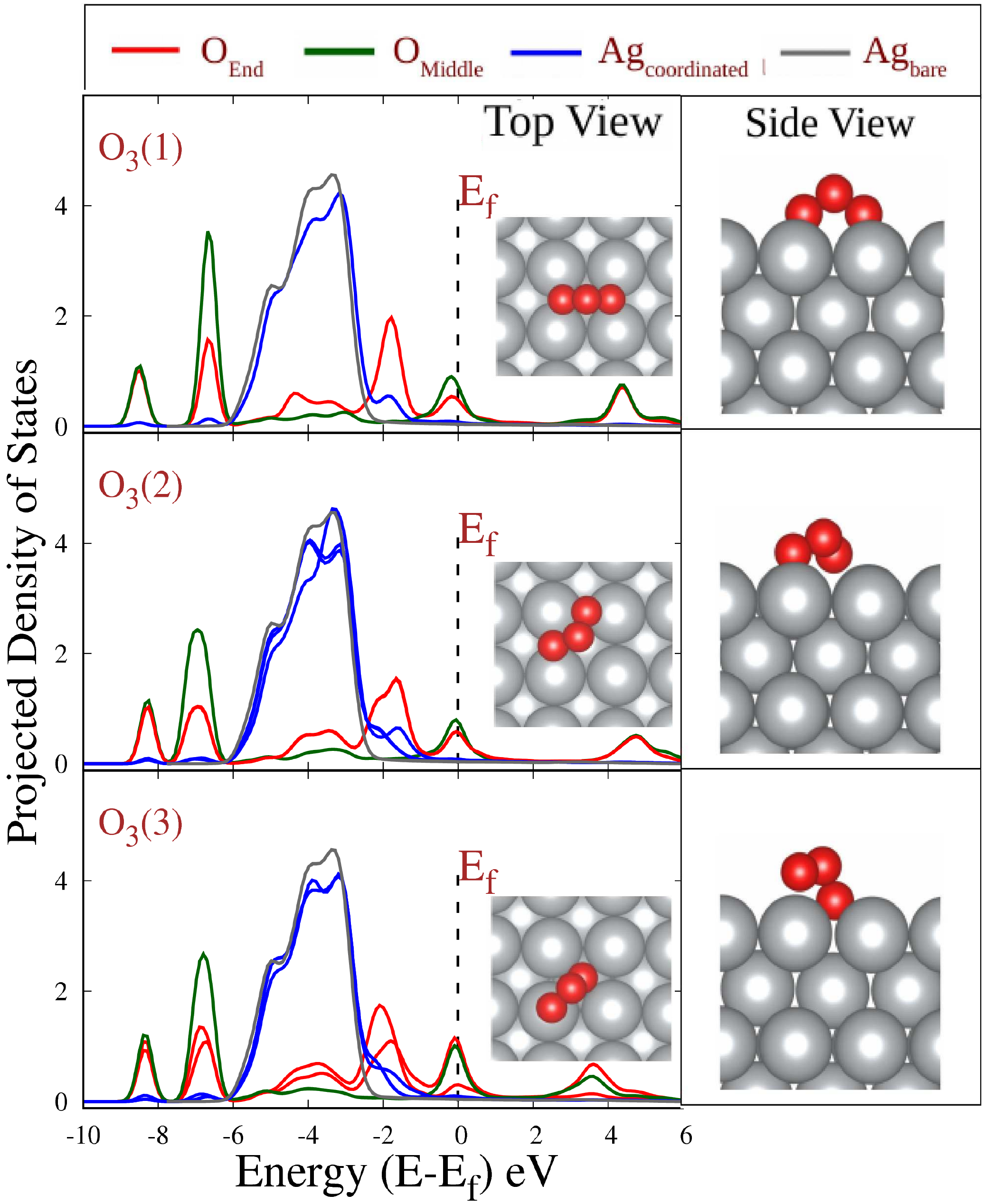}}
  \end{minipage}
  \begin{minipage}[b]{0.47\linewidth}
   \centering
   \subfloat[]{\label{fig4:b} \includegraphics[width=\textwidth]{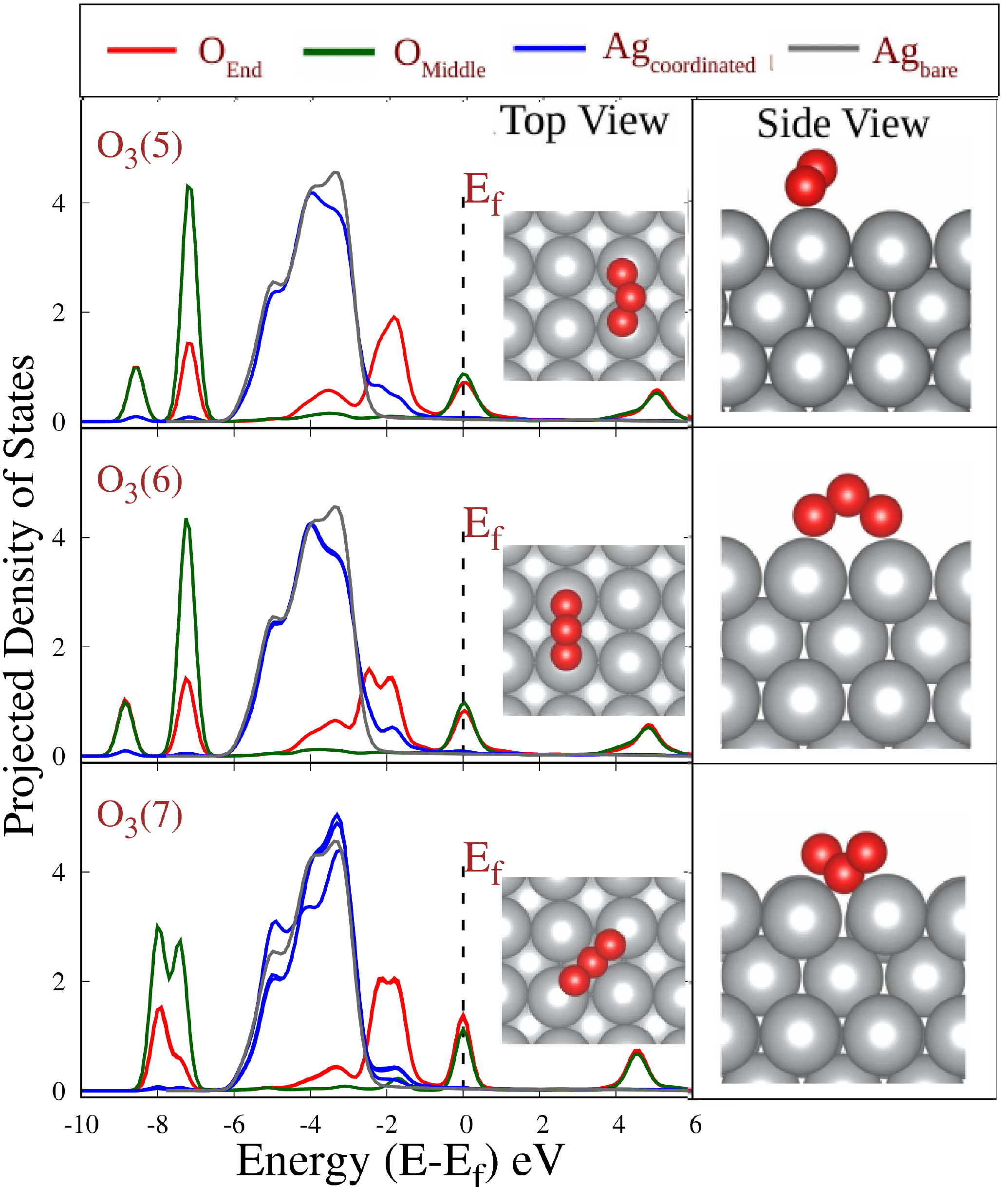}}
\end{minipage}
  \caption{\small {Site specific {\it pdos} for O$_{3}$ conf\mbox{}igurations at $0.75$ ML along with top and side view of 
stable conf\mbox{}igurations.
 2p states of O$_{mid}$ are shown in green and 2p states of O$_{end}$ atoms are shown in red. Availability of empty states 
near E$_f$ are observed in all the configurations
}}
  \label{fig4}
\end{figure}

\begingroup
\setlength{\tabcolsep}{8pt}
\renewcommand{\arraystretch}{0.9}
\begin{table}[h!]
\centering
\begin{tabular}{c c c c c c c}
Sr. & Conf\mbox{}iguration & $\Delta$E  & Ag-O           & Ag-O            & O-O             & L\"owdin Charge \\
No. & Description   &       (eV) & coordination  & bond length(\AA) & bond length(\AA) & on O          \\
\hline

1. & \textcolor{blue}{O$_{2}$O(1)} & \textcolor{blue}{0.000}&\textcolor{blue}{O$_{2}$=2}&\textcolor{blue}{2.38}&\textcolor{blue}{1.33}&
\textcolor{blue}{O$_{2}$=6.08} \\
   &                   	 &		     &\textcolor{blue}{O=4}      & \textcolor{blue}{2.21}& \textcolor{blue}{-} & \textcolor{blue}{O=6.50} \\ 

2. & \textcolor{blue}{O$_{2}$O(2)} &\textcolor{blue}{0.043} &\textcolor{blue}{O$_{2}$=2} &\textcolor{blue}{2.35} & \textcolor{blue}{1.35} &\textcolor{blue}{O$_{2}$=6.11} \\
   &                               &              & \textcolor{blue}{O=4}  & \textcolor{blue}{-} & \textcolor{blue}{2.22}& \textcolor{blue}{O=6.48} \\

3. & \textcolor{green!30!black}{O$_{3}$(1)} &\textcolor{green!30!black}{0.083} &\textcolor{green!30!black}{O$_{End}$=2}&\textcolor{green!30!black}{2.28}&\textcolor{green!30!black}{1.39} &\textcolor{green!30!black}{O$_{End}$=6.22} \\
   &                                  &              &\textcolor{green!30!black}{O$_{Mid}$=0} & \textcolor{green!30!black}{-} &\textcolor{green!30!black}{-} &\setlength{\fboxrule}{0.7pt}\fbox{{\textcolor{green!30!black}{O$_{Mid}$=5.85}}}\\  

4. & \textcolor{green!30!black}{O$_{3}$(2)} &\textcolor{green!30!black}{0.229} &\textcolor{green!30!black}{O$_{End}$=2} &\textcolor{green!30!black}{2.27-2.42}&\textcolor{green!30!black}{1.37}          &\textcolor{green!30!black}{O$_{End}$=6.22} \\
   &                                  &         &\textcolor{green!30!black}{O$_{Mid}$=0} & \textcolor{green!30!black}{-} &\textcolor{green!30!black}{-}&\setlength{\fboxrule}{0.7pt}\fbox{\textcolor{green!30!black}{O$_{Mid}$=5.82}} \\

5. & \textcolor{blue}{O$_{2}$O(3)}&\textcolor{blue}{0.306} &\textcolor{blue}{O$_{2}$=0} &\textcolor{blue}{-} &\textcolor{blue}{1.28} &\textcolor{blue}{O$_{2}$=6.03} \\
   &                               &  & \textcolor{blue}{O=4} & \textcolor{blue}{2.23} &\textcolor{blue}{-} &\textcolor{blue}{O=6.50}\\

6. & \textcolor{green!30!black}{O$_{3}$(3)}  &\textcolor{green!30!black}{0.340} &\textcolor{green!30!black}{O$_{End1}$=2} &\textcolor{green!30!black}{2.29} &\textcolor{green!30!black}{1.32-1.47} &\textcolor{green!30!black}{O$_{End1}$=6.27} \\
   &          	 &   &\textcolor{green!30!black}{O$_{End2}$=1} &\textcolor{green!30!black}{2.40} & \textcolor{green!30!black}{-} &\textcolor{green!30!black}{O$_{End2}$=6.16} \\
   &             &   & \textcolor{green!30!black}{O$_{Mid}$=0} & \textcolor{green!30!black}{-} & \textcolor{green!30!black}{-} &\setlength{\fboxrule}{0.7pt}\fbox{\textcolor{green!30!black}{O$_{Mid}$=5.83}} \\	

7. & \textcolor{green!30!black}{O$_{3}$(4)}        &\textcolor{green!30!black}{0.343}	&\textcolor{green!30!black}{O$_{End1}$=3} &\textcolor{green!30!black}{2.39-2.43} &\textcolor{green!30!black}{1.32-1.48} &\textcolor{green!30!black}{O$_{End1}$=6.28} \\
   &	&     &\textcolor{green!30!black}{O$_{End2}$=1} &\textcolor{green!30!black}{2.30} &\textcolor{green!30!black}{-} &\textcolor{green!30!black}{O$_{End2}$=6.16} \\
   &	&     &\textcolor{green!30!black}{O$_{Mid}$=0} & \textcolor{green!30!black}{-} & \textcolor{green!30!black}{-} &\setlength{\fboxrule}{0.7pt}\fbox{\textcolor{green!30!black}{O$_{Mid}$=5.83}} \\

8. & \textcolor{green!30!black}{O$_{3}$(5)}        &\textcolor{green!30!black}{0.548} &\textcolor{green!30!black}{O$_{End}$=1} &\textcolor{green!30!black}{2.24} &\textcolor{green!30!black}{1.34} &\textcolor{green!30!black}{O$_{End}$=6.19} \\
   &	&	 &\textcolor{green!30!black}{O$_{Mid}$=0} & \textcolor{green!30!black}{-} & \textcolor{green!30!black}{-} &\setlength{\fboxrule}{0.7pt}\fbox{\textcolor{green!30!black}{O$_{Mid}$=5.75}} \\

9. & \textcolor{green!30!black}{O$_{3}$(6)}       &\textcolor{green!30!black}{0.558} &\textcolor{green!30!black}{O$_{End}$=1} &\textcolor{green!30!black}{2.24} &\textcolor{green!30!black}{1.34} &\textcolor{green!30!black}{O$_{End}$=6.17} \\
   &	& &\textcolor{green!30!black}{O$_{Mid}$=0} & \textcolor{green!30!black}{-} & \textcolor{green!30!black}{-} &\setlength{\fboxrule}{0.7pt}\fbox{\textcolor{green!30!black}{O$_{Mid}$=5.73}} \\

10. & \textcolor{blue}{O$_{2}$O(4)} &\textcolor{blue}{1.010}  &\textcolor{blue}{O$_{2}$=1} &\textcolor{blue}{2.28} & \textcolor{blue}{1.27} &\setlength{\fboxrule}{0.7pt}\fbox{\textcolor{blue}{O$_{2}$=5.98}} \\
    & & &\textcolor{blue}{O=2}  &\textcolor{blue}{2.03} &\textcolor{blue}{-} &\textcolor{blue}{O=6.43} \\

11. & \textcolor{blue}{O$_{2}$O(5)} &\textcolor{blue}{1.070} &\textcolor{blue}{O$_{2}$=1} &\textcolor{blue}{2.31} &\textcolor{blue}{1.28} & \setlength{\fboxrule}{0.7pt}\fbox{\textcolor{blue}{O$_{2}$=5.99}} \\
    & &	  &\textcolor{blue}{O=2}  &\textcolor{blue}{2.04} &\textcolor{blue}{-} &\textcolor{blue}{O=6.42} \\

12. & \textcolor{green!30!black}{O$_{3}$(7)} &\textcolor{green!30!black}{1.441} &\textcolor{green!30!black}{O$_{End}$=1} &\textcolor{green!30!black}{2.43} &\textcolor{green!30!black}{1.36}          &\textcolor{green!30!black}{O$_{End}$=6.20} \\
    &	&  &\textcolor{green!30!black}{O$_{Mid}$=0} & \textcolor{green!30!black}{-} & \textcolor{green!30!black}{-} &\setlength{\fboxrule}{0.7pt}\fbox{\textcolor{green!30!black}{O$_{Mid}$=5.77}} \\

\hline
\end{tabular}
\caption{\small {Relative energy ($\Delta$E), Ag-O coordination along with bond length,
and L\"owdin charges on oxygen for dif\mbox{}ferent positions of oxygen at 0.75 ML coverage.
Conf\mbox{}igurations like combination of atomic and molecular O are shown in blue (O$_{2}$O(1) to O$_{2}$O(5))
and triatomic molecular conf\mbox{}igurations (O$_{3}$(1) to O$_{3}$(7) in green color.}}
\label{tab3}
\end{table}
\endgroup

The conf\mbox{}iguration (O$_{2}$O(1)) in which both atomic and molecular
oxygen adsorbed at hollow site has the lowest energy. 
However, if we consider ten most stable configurations they are dominated by adsorption of triatomic molecular oxygen as evident
from Table\ref{tab3}
Keen observation of Fig.\ \ref{fig3} reveals that {\it pdos} of O$_{2}$O conf\mbox{}igurations coincide with combined {\it pdos} 
of atomic O (shown in Fig.\ \ref{fig1}) and molecularly adsorbed oxygen (shown in Fig.\ \ref{fig2}). 
This suggests that, these entities do not af\mbox{}fect the chemical nature of each other. L\"owdin charges on oxygen 
for O$_{2}$O(4) and O$_{2}$O(5) show marginal positive charge; but empty states which are indispensable for direct 
epoxidation are not present beyond E$_{f}$ in nearby locus. Thus, we precluded these conf\mbox{}igurations as 
candidates for electrophilic oxygen.

Fig.\ \ref{fig4} represents top and side view of six out of seven O$_{3}$ conf\mbox{}igurations along with 
their respective {\it pdos}.\footnote{The one with similar {\it pdos} is not included.}
Adsorption of three oxygen as O$_{3}$ on Ag surface leads to end oxygens coordinating with Ag. 
As Table\ \ref{tab3} suggests that middle oxygen(O$_{mid}$) of all O$_{3}$ conf\mbox{}igurations 
possess signif\mbox{}icant positive charge (highlighted in Table\ref{tab3}). Further, as seen in Fig.\ \ref{fig4} 
all these configurations with adsorbed O$_3$ show 
presence of empty states beyond E$_{f}$ in energy range $3.5$-$5.5$ eV which is a characteristic 
feature of electron def\mbox{}icient species. 

If direct epoxidation by means of electrophilic attack of oxygen on ethene is the main mode of industrial epoxidation 
then such O$_{ele}$ must feature characteristics of an electrophile. In case of ethene which has $\pi$ orbitals at 
E$_{f}$ and $\pi^*$ orbital at $6$ eV, O$_{ele}$ must possess empty states matching energy range of ethene 
$\pi$.\cite{kokalj_2002} This will facilitate overlap among these orbitals into which $\pi$ electrons can be 
transferred. In case of O$_{3}$ species, this could be further facilitated by positive charge on middle oxygen. 
Again presence of weak O-O bonds in this species will further help dissociation of middle oxygen from O$_{3}$ moiety. 
Also, a notable fact is that this state arises out of O-2p interactions alone and have no contribution from 
Ag-4d or Ag-5s orbitals, this is contradictory to the presumption that O$_{ele}$ must have signif\mbox{}icant 
overlap with Ag-4d or Ag-5s.\cite{jones_insights_2015,bukhtiyarov_2001,bao_interaction_1996} 

\begin{figure}[h]
  \centering
    \includegraphics[width=0.60\textwidth]{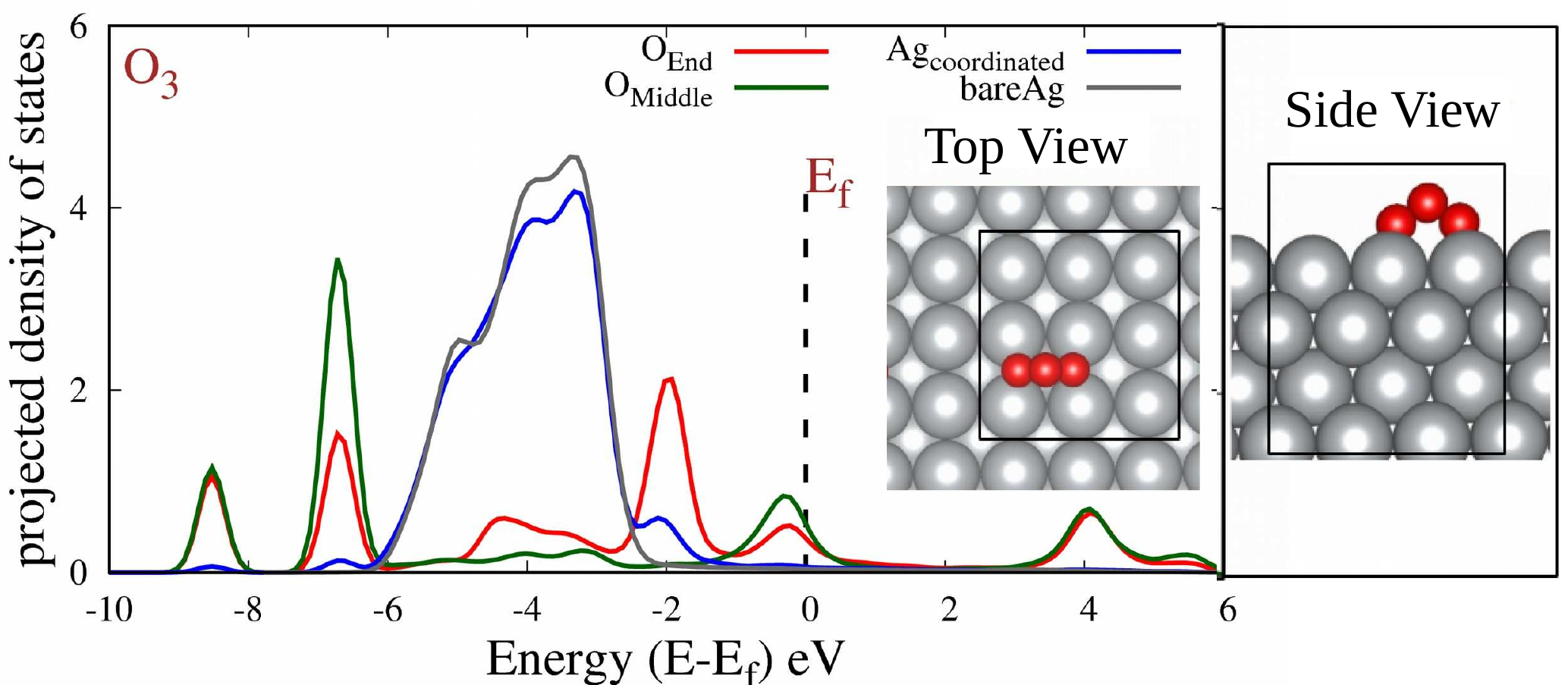}
  \caption{Site specific {\it pdos} for O$_{3}$ conf\mbox{}iguration on a 3x3 supercell along with top 
	and side view of the conf\mbox{}iguration.  2p states of O$_{mid}$ are shown in green and 2p 
	states of O$_{end}$ atoms are shown in red. Availability of empty states near E$_f$ are observed 
	demonstrating that it is not an artifact of a smaller slab}
  \label{fig5}
\end{figure}

To demonstrate that O$_3$ is not an artifact of a smaller slab that has been modeled here, 
we have also computed {\it pdos} of O$_3$ on a larger supercell. Fig.\ \ref{fig5} shows 
the site dependent {\it pdos} for O$_3$ on a 3x3 supercell. O$_3$ exhibits similar features 
in {\it pdos} demonstrating that indeed middle oxygen in O$_3$ moiety bares signatures of 
electrophilic oxygen. The L\"owdin charge on the middle oxygen is $5.896$. Thus
it fulfills both the conditions of an electrophile.

\begin{figure}[h]
  \centering
    \includegraphics[width=0.75\textwidth]{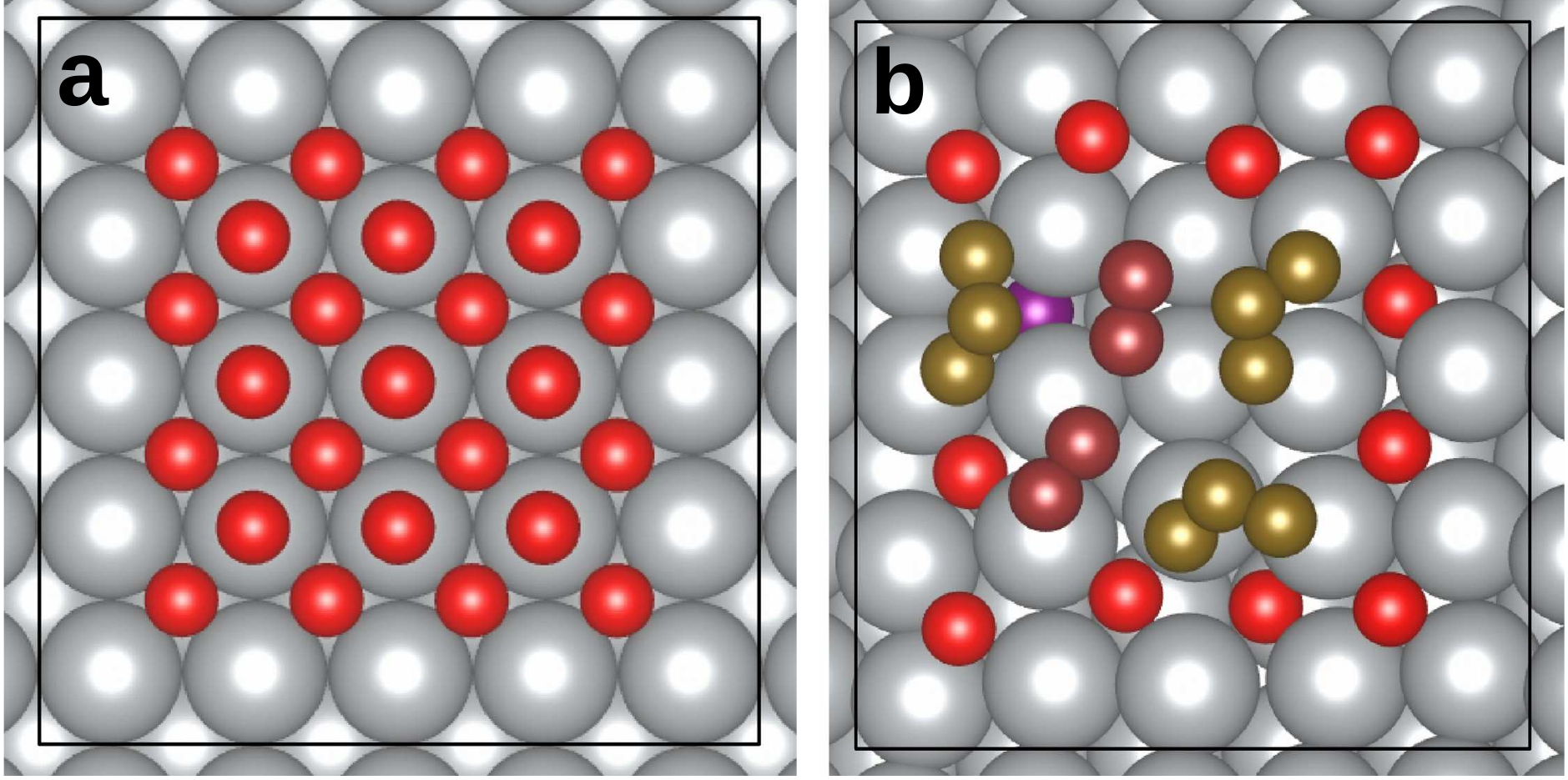}
  \caption{\small {For 5X5 Ag 100 surface shown in a box adsorbed with 25 oxygen a) initial configuration (top view) in which 16 oxygens placed at 4FH
and 9 oxygens are on 1FT positions, b) final conf\mbox{}iguration (top view) shows presence of three types of oxygens viz,. atomic oxygen O (red), 
diatomic molecular oxygen O$_{2}$ (brown) and triatomic molecular oxygen O$_{3}$ (golden rod)} }
  \label{fig6}
\end{figure}
\begin{figure}[h]
  \centering
	\includegraphics[width=0.75\textwidth]{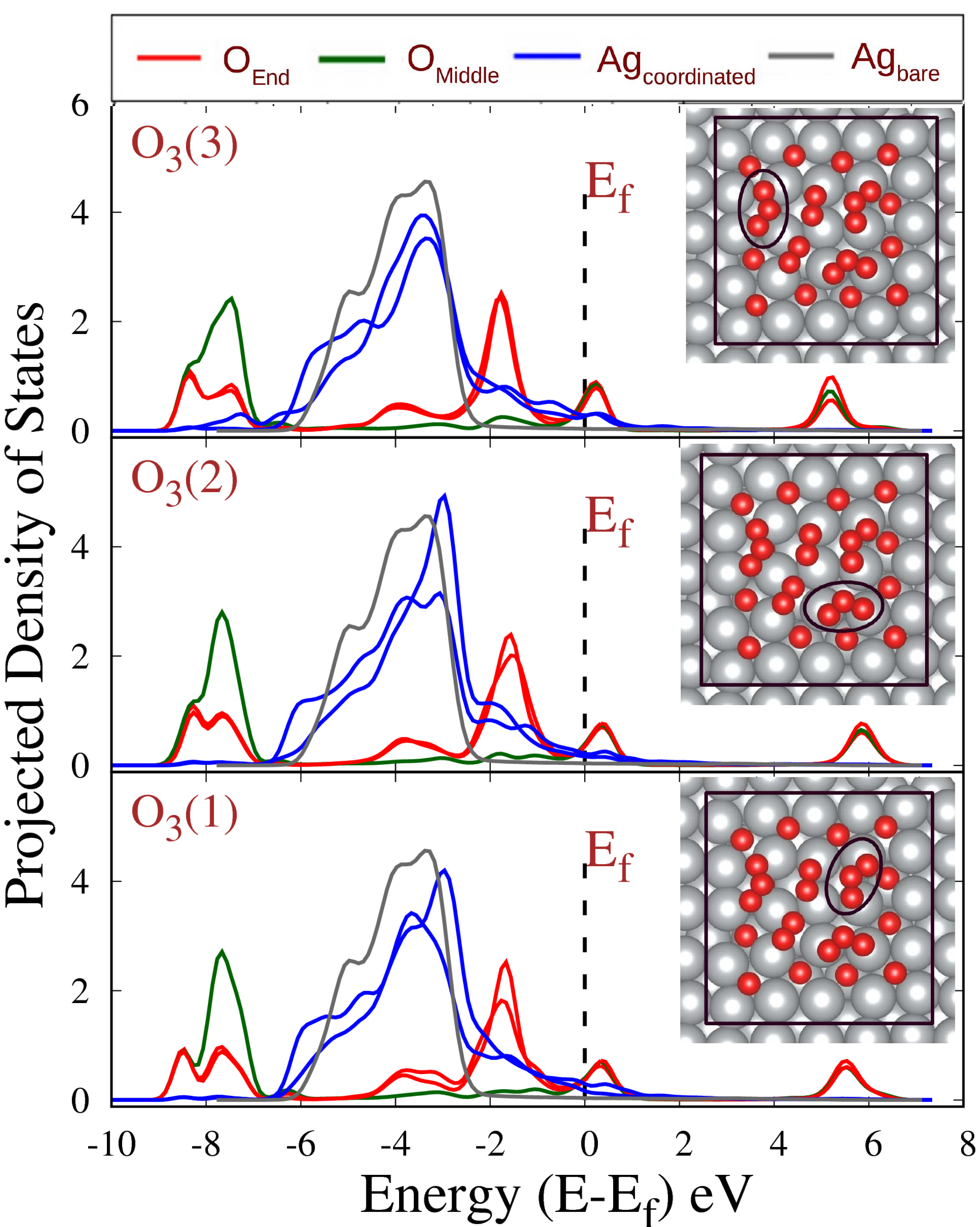}
  \caption{Site specific {\it pdos} for O$_{3}$ conf\mbox{}igurations on a 5x5 supercell along with top
        and side view of the conf\mbox{}igurations.  2p states of O$_{mid}$ are shown in green and 2p
	states of O$_{end}$ atoms are shown in red. Availability of empty states near E$_{f}$ confirms that
	O$_{mid}$ is electrophilic in nature}
  \label{fig7}
\end{figure}

However, prevalence of the O$_{3}$ will be much more at elevated pressures. To demonstrate this we have also 
optimized a 5x5 supercell at 1 ML coverage. The initial configuration along with optimized structure are
shown in Fig\ \ref{fig6}. The initial configuration (see Fig.\ \ref{fig6}-a) consists of
sixteen oxygen atoms at hollow positions and nine oxygen atoms at on top positions resulting into 1 ML coverage. 
The relaxed structure is shown
in Fig.\ \ref{fig6}-b. Different oxygen moieties are shown in different colors to aid an  eye. 
This also brings out the complexity of the problem at hand.
The optimized structure consists of three O$_{3}$ moieties, two O$_{2}$ molecules with extended O-O bondlength. 
There are
few oxygen atoms at hollow sites as well as few oxygen atoms have escalated to subsurface site.
The figure also brings out
the extensive surface reconstruction that took place upon relaxation. A point to be noted here is 
limitation of our 2x2 slab to bring out such surface reconstructions. However, the most important feature is
all O$_{3}$ moieties observed in this structure bare both signatures of
electrophilic oxygen. Only three middle oxygens in the three O$_{3}$ moieties have partially positive L\"{o}wdin 
charge. Rest of all oxygens are partially negatively charged although the magnitude of the charge varies
based on their actual position. Further, the site dependent {\it pdos} for all the O$_{3}$ moieties (shown in 
Fig.\ \ref{fig7}) exhibit empty states above E$_f$. The {\it pdos} for O$_{2}$ and oxygen at hollow position 
are also similar to what we have shown in our 2x2 model except minor modifications due to surface reconstructions.

Thus, our simulations convincingly brings out signatures of electrophilic oxygen and scenario where this
moiety will be present.
At this point, an important question should be asked is, then why it was not detected so far?
As our investigations show, O$_{3}$ will be more probable at higher pressures as well as temperatures.
Although, the industrial catalyst for EtO works at higher pressure and temperature, most of the surface 
science studies are carried out at low pressures or very low pressures. And there is a pressure gap between
the surface science experiments and working conditions of an industrial catalyst. We believe this is the most
probable reason for not so successful search for the electrophilic oxygen.
Few caveats should be noted when we are closing our discussions. First, scope of this work is limited, i.e. 
to search for the electrophilic oxygen. The formation of EtO on Ag surfaces have many aspects which are not 
being touched upon in this work and there are many open questions still exist and awaits explanations.

\section{\label{sec:concl}Conclusions} 
Ethylene epoxidation is one of the most investigated reactions in heterogeneous catalysis. Two types 
of oxygen species, electrophilic and nucleophilic oxygen were envisaged. Although, many experiments and
simulations could bring out signatures of nucleophilic oxygen, so far electrophilic oxygen remains
a mystery. In the present work, we investigate interaction between Ag(100) surface and oxygen as a function of 
monolayer concentration.  Our extensive investigations reveal that at 0.25ML and 0.5ML
the two signatures associated with electrophilic oxygen species viz. positive charge on oxygen and 
empty states near Fermi level are missing and so we can not consider these atomically or molecularly adsorbed
oxygen as an electrophilic oxygen required for direct epoxidation. On the other hand, adsorbed O$_3$ species 
at 0.75 ML concentration bares signatures relevant for electrophilic oxygen. Middle oxygen in {\it all} 
adsorbed O$_3$ have significant positive charge as evident from their L\"{o}wdin charges as well as {\it all} these
configurations have empty states in between 3.5 -- 5.5 eV required to accept $\pi$ electrons from ethene to form 
EtO. Thus, our investigations bring out situations corresponding to electrophilic oxygen required for 
direct epoxidation.

\section{Acknowledgements}

We are thankful to Dr. C. S. Gopinath for many fruitful discussions. CSIR-4PI is gratefully acknowledged for 
the computational facility. KJ and NK acknowledge DST (EMR/2016/000591) for partial financial support. 
SP acknowledges CSIR for research fellowship.

\bibliography{Epoxidation}

\end{document}